\documentclass[sigconf,screen]{acmart}
\newcommand{\name}{\mbox{ProfOlaf}}
\newcommand{\TODO}[1]{\textcolor{red}{#1}\GenericWarning{}{LaTeX Warning: TODO: #1}}\newcommand\todo\TODO

\usepackage[table]{xcolor} %
\usepackage{multirow}
\usepackage{booktabs}
\usepackage{threeparttable}

\AtBeginDocument{%
  }

\setcopyright{acmlicensed}
\copyrightyear{2018}
\acmYear{2018}
\acmDOI{XXXXXXX.XXXXXXX}
 \acmConference[Conference acronym 'XX]{Make sure to enter the correct
  conference title from your rights confirmation email}{June 03--05,
  2018}{Woodstock, NY}

\acmBooktitle{Companion Proceedings of the 34th ACM Symposium on the Foundations of Software Engineering (FSE '26), June 5--9, 2026, Montreal, Canada}
\acmISBN{978-1-4503-XXXX-X/2018/06}

\begin{document}

\title{ProfOlaf: Semi-Automated Tool for Systematic Literature Reviews}

\author{Martim Afonso}
\authornotemark[1]
\orcid{https://orcid.org/0009-0002-9271-9276}
\affiliation{%
  \institution{INESC-ID \& IST, University of Lisbon}
  \city{Lisbon}
  \country{Portugal}
}
\affiliation{%
  \institution{Politecnico di Torino}
\city{Turin}
  \country{Italy}
}
\email{martim.afonso@tecnico.ulisboa.pt}

\author{Nuno Saavedra}
\authornote{Both authors contributed equally to this research.}
\orcid{https://orcid.org/0000-0003-4148-5991}
\affiliation{%
  \institution{INESC-ID \& IST, University of Lisbon}
  \city{Lisbon}
  \country{Portugal}
}
\email{nuno.saavedra@tecnico.ulisboa.pt}

\author{Bruno Lourenço}
\orcid{https://orcid.org/0009-0005-9077-2040}
\affiliation{%
  \institution{INESC-ID, IST \& CINAV, University of Lisbon \& Portuguese Naval Academy}
  \city{Lisbon}
  \country{Portugal}
}
\email{bruno.horta.lourenco@tecnico.ulisboa.pt}

\author{Alexandra Mendes}
\orcid{0000-0001-8060-5920}
\affiliation{%
 \institution{INESC TEC, Faculty of Engineering, University of Porto}
 \city{Porto}
 \country{Portugal}}
\email{alexandra@archimendes.com}

\author{João F. Ferreira}
\orcid{0000-0002-6612-9013}
\affiliation{%
  \institution{INESC-ID \& Faculty of Engineering, University of Porto}
  \city{Porto}
  \country{Portugal}
}
\email{joao@joaoff.com}

\renewcommand{\shortauthors}{Afonso et al.}

\begin{abstract}
  Systematic reviews and mapping studies are critical to synthesize research, identify gaps, and guide future work, but are often labor-intensive and time-consuming. Existing tools provide partial support for specific steps, leaving much of the process manual and error-prone. We present \name, a semi-automated tool designed to streamline systematic reviews while maintaining methodological rigor. \name\ supports iterative snowballing for article collection with human-in-the-loop filtering and uses large language models to help select articles, extract key topics, and answer queries about the content of articles. By combining automation with guided manual effort, \name\ enhances the efficiency, quality, and reproducibility of systematic reviews across research fields. \name\ can be used both as a CLI tool and in web application format. A video demonstrating \name\ is available at: \url{https://youtu.be/R-gY4dJlN3s}
\end{abstract}



\received{20 February 2007}
\received[revised]{12 March 2009}
\received[accepted]{5 June 2009}

\maketitle

\section{Introduction}
Systematic literature reviews and mapping studies play an essential role across research fields, as they organize and synthesize existing knowledge, providing a structured overview that highlights established findings, identifies gaps, and indicates promising directions for future research.
Unlike other forms of literature reviews, systematic reviews hold particular scientific value because they follow a transparent, well-defined, and unbiased methodology~\cite{brereton2007lessons}.

Despite their value, conducting systematic reviews is labor-intensive and time-consuming~\cite{zhang2013systematic, van2021automation}. The process typically requires conducting broad and comprehensive searches in multiple academic databases, followed by careful filtering of large volumes of articles against predefined inclusion and exclusion criteria. After the collection is complete, the reviewers must also inspect and analyze the selected studies in depth, identify recurring research topics, and formulate answers to predefined research questions. Each of these steps demands sustained effort and precision.

Several methods and tools have been proposed to automate or support various steps of the systematic review process~\cite{fabiano2024optimize, van2021automation}.
Existing solutions, such as reference managers or search engines, offer only partial assistance, leaving key tasks manual, time-consuming, and error-prone, reducing efficiency and reproducibility~\cite{keele2007guidelines}. More advanced approaches, such as AI-based tools, have shown promising results~\cite{fabiano2024optimize}, but remain fragmented. Although these tools exist for individual steps of the %
process, no solution provides comprehensive support at all stages. %
Furthermore, achieving a sensible balance between manual effort and tool assistance is essential to ensure that precision is not compromised in the pursuit of reducing effort.

In response to these requirements, we propose \name, a semi-automated tool designed to support and streamline the review process. \name\ implements a structured methodology that adheres to established guidelines for systematic reviews. The methodology was designed with the goal of reducing human effort to the greatest extent possible. For the collection phase, \name\ uses an iterative snowballing process to collect relevant articles, each iteration involving human-led filtering supported by our tool and optionally assisted by large language models (LLMs) as secondary reviewers.
In the analysis phase, \name\ integrates LLMs to ease the analysis of the collected articles, enabling users to extract key topics from each article and query the model regarding its contents. 

\begin{figure*}
    \centering
    \includegraphics[width=0.67\linewidth]{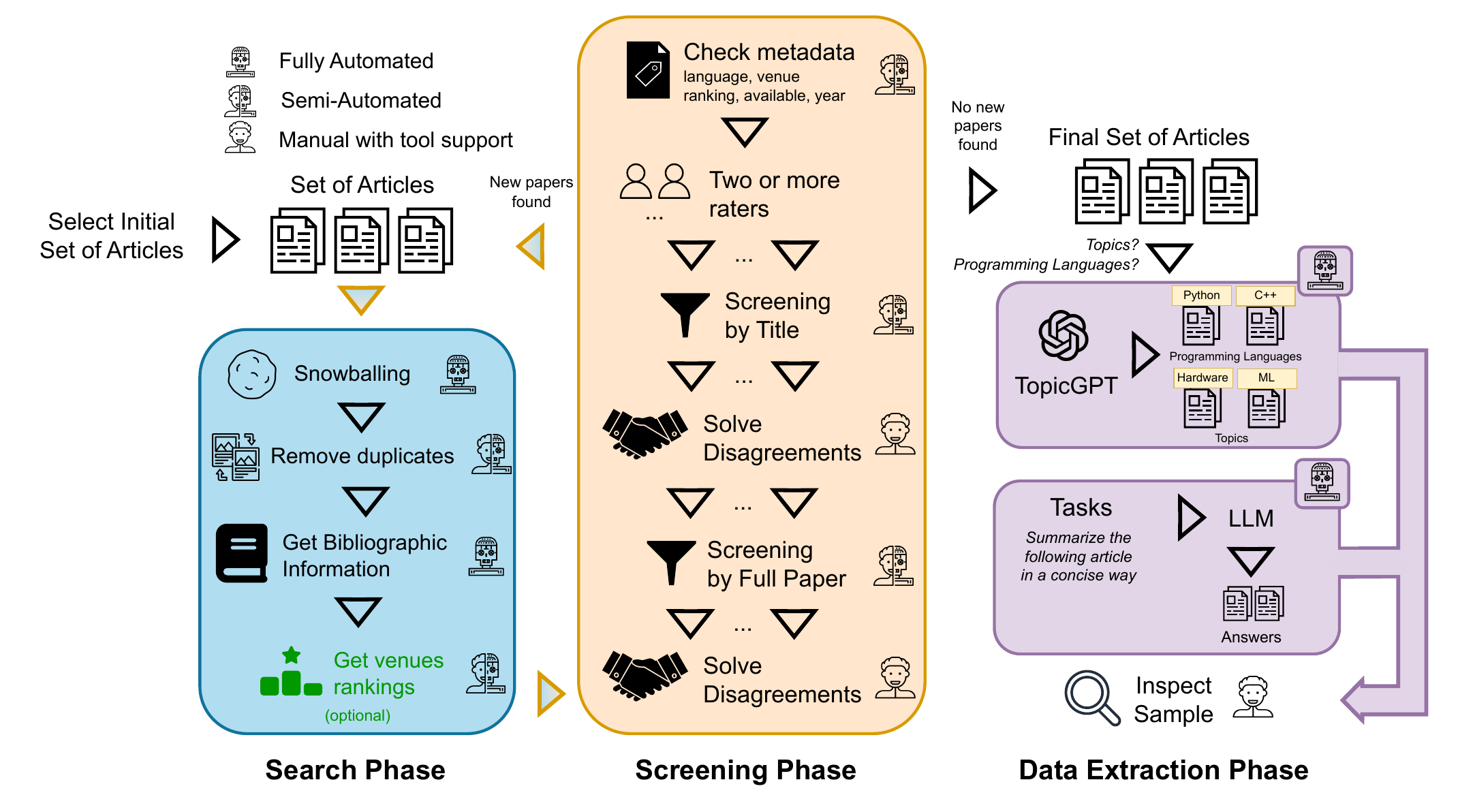}
    \caption{Overview of the ProfOlaf methodology.}
    \label{fig:search_phase}
\end{figure*}

\name\ assists researchers in conducting systematic reviews, improving both their quality and scalability. It is open source and available at \url{https://github.com/sr-lab/ProfOlaf} or at \url{https://doi.org/10.5281/zenodo.18386217} (as a container), together with the experimental artifacts used in our evaluation. \name\ is provided as both a command-line interface and a web application. 

\section{\name\ Overview}

The \name\ methodology is illustrated in Figure~\ref{fig:search_phase}.
This section provides a detailed explanation of each step of the pipeline.

\subsection{Initial Setup}\label{initial_setup}
To setup a new search with \name\, the user starts by defining a file containing an initial set of article titles. These articles are previously selected by the user and can be obtained through, for instance, a previous literature review on the topic being handled. The tool receives this file and generates a database that stores all the metadata of the articles as well as the intermediate states of each article during the search.

\subsection{Snowballing}\label{snowballing}
We chose snowballing as our search method, as prior studies show it performs as well as or better than database searches~\cite{badampudi2015experiences, jalali2012systematic}.

After populating the database with the initial set of articles, \name\ retrieves either the citations (\emph{forward snowballing}), the references (\emph{backward snowballing}), or both, for each article. The tool then compiles the bibliographic information for the entire set.

\name\ obtains article data from Google Scholar~\cite{google_scholar}, Semantic Scholar~\cite{semantic_scholar}, and DBLP~\cite{dblp}. However, the tool has been designed to ease the integration of additional search sources.
As multiple versions of the same paper may exist under slightly different titles, this stage also includes a mechanism to identify and present these versions, allowing the user to select which one to keep.

\subsection{Metadata Screening}\label{metadata_filtering}
Subsequently, \name\ filters the retrieved articles using the metadata collected in the preceding step. Filtering criteria include venue ranking, publication year, and language, all of which are optional.
Kitchenham et al. have recommended such practical criteria to refine the selection of articles~\cite{kitchenham2007guidelines}.

When the venue-ranking filter is applied, the user must first execute the step that assigns a ranking to all the venues present in the collected articles.
To help the user with this manual task, \name\ first uses cosine similarity to find venues previously ranked that are similar to the one being classified. Then, the tool searches for the venue in venue ranking databases and presents the top results for each database. For each result, \name\ outputs the title of the venue, its registered ranking, and the cosine similarity score of the search. Currently, \name\ searches both in the SCImago and CORE databases, but this functionality is easily extendable to other venue ranking databases.

\subsection{Article Screening}\label{manual_filtering}
\name\ adopts the approach outlined by Wohlin~\cite{wohlin2014guidelines}, which recommends screening articles progressively: first the title, then the abstract, and finally the full text.

As such, in this phase, the user is prompted to further refine the results, starting with removing irrelevant articles by title and then by their full content. For this, \name\ presents, for each article, its title as well as a \textit{url} where the user can access it.
After each step, the tool can be used to identify and display discrepancies between the user and other reviewers' assessments. The reviewers must then discuss these disagreements and reach a consensus. The result of this discussion can then be inserted into the tool.

In addition, the tool offers the option of using an LLM as an auxiliary rater to provide additional input during the discussion. Section~\ref{sec:auto_screening_eval} presents an evaluation of this component.

With both screening phases completed, the resulting set of articles is used as input to the snowballing phase (Section~\ref{snowballing}) to start a new iteration. If an iteration does not produce new results, \name\ consolidates the findings from all iterations into a single file.

\subsection{Topic Modeling and Task Assistant}

After the collection process is completed, the user can use \name\ to ease the manual process of article analysis through the use of LLMs. For this, our tool provides functionality for downloading all selected articles. Then the content of the PDFs is parsed and can be provided to two analysis modules: topic modeling and our task assistant. The extracted information from both modules should be manually verified by users, either in its entirety or by sampling.

The topic modeling module uses TopicGPT~\cite{pham2024topicgpt} to gather different topics from the collection and cluster them according to those topics.
TopicGPT is a prompt-based framework that uses LLMs to generate interpretable topics with natural language labels and descriptions. It enables users to group papers for deeper analysis, offering greater transparency and verifiability than traditional bag-of-words approaches. 
This functionality can also be useful for addressing specific research questions with closed-form answers, such as \emph{``Which programming language is being considered?''}.

The task assistant module enables users to submit an article to a public-access LLM and request tasks such as key information extraction or concise, query-tailored summaries.

\section{Evaluation}

To evaluate \name, we conducted a small illustrative systematic review using Ramos et al.~\cite{ramos2411large} as the seed article. This paper was selected for its Best Paper Award at LLM4Code 2025 and our familiarity with its topic.

\subsection{Search and Screening Phase}

The screening phase was carried out over seven iterations by two human raters, corresponding to the second and third authors.
The following inclusion criteria were applied: (1) related to Machine Learning for Code; (2) written in English; (3) publicly available; and (4) published in a venue ranked by CORE or Scimago.

The results of all iterations are summarized in Table~\ref{tab:iterations}.
We calculate the efficiency measure for systematic literature reviews, a metric used by Wohlin~\cite{wohlin2014guidelines} to evaluate the amount of noise in the search. 
Efficiency is calculated as the number of included articles relative to the total number of candidate articles examined.
The efficiency by iteration and the final efficiency are both represented in Table~\ref{tab:iterations}.
After screening, we obtained 108 articles, including the seed article by Ramos et al.~\cite{ramos2411large}.
 
\subsection{Automated Screening}\label{sec:auto_screening_eval}
We used \name, configured to use \emph{gpt5.2}, to perform both screening stages using a concise topic description along with inclusion and exclusion criteria. To evaluate its performance, we compared the model's decisions against the final consensus reached by the two original raters. We also introduced an additional independent rater (fifth author) to screen the articles, allowing us to assess how the LLM's performance compared to that of an individual human. Table~\ref{tab:screening-eval} presents these findings. The screening process encompassed 183 articles at the title level and 125 articles at the full-text level.

We can observe that the LLM had a performance comparable to that of a human, with slightly greater precision, but lower recall, which indicates that the LLM tended to be more conservative in its screening decisions, favoring correctness over coverage.

\begin{table}
\centering
\footnotesize
\caption{Snowballing process results across iterations.}
\label{tab:iterations}
\begin{threeparttable}
\begin{tabular}{ccccc}
\toprule
\textbf{I} & \textbf{Retrieved} & \textbf{Rejected (Meta. + Screen.)} & \textbf{Approved} & \textbf{Efficiency} \\
\midrule
1 & 19  & 13 + 1   & 5  & 0.26 \\
2 & 100 & 63 + 7   & 29 & 0.29 \\
3 & 227 & 158 + 47 & 22 & 0.09 \\
4 & 111 & 84 + 9   & 18 & 0.16 \\
5 & 100 & 72 + 3   & 24 & 0.24 \\
6 & 433 & 414 + 9  & 10 & 0.02 \\
7 & 19  & 19 + 0   & 0  & 0.00 \\
\midrule
Total & 1009 & 823 + 76 & 108 & 0.11 \\
\bottomrule
\end{tabular}

\begin{tablenotes}\tiny
\item \textit{Notes:} I = iteration; Meta. = metadata; Screen. = Screening.
\end{tablenotes}
\end{threeparttable}
\end{table}

\begin{table}
\centering
\footnotesize
\caption{Automated screening evaluation.}
\label{tab:screening-eval}
\begin{tabular}{llcccc}
\toprule
\textbf{Rater} & \textbf{Screening} & \textbf{Accuracy} & \textbf{Precision} & \textbf{Recall} & \textbf{F1 Score} \\
\midrule
Human & Title        & 0.8361 & 0.8613 & \textbf{0.9147} & \textbf{0.8872} \\
      & Full Content & 0.8760 & 0.9314 & \textbf{0.9223} & 0.9268 \\
\addlinespace
LLM   & Title        & \textbf{0.8415} & \textbf{0.9098} & 0.8605 & 0.8845 \\
      & Full Content & \textbf{0.8800} & \textbf{0.9417} & 0.9151 & \textbf{0.9282} \\
\bottomrule
\end{tabular}
\end{table}

\subsection{Data Extraction Phase}

We selected two tasks for the Topic Modeling module: identifying the programming languages explored and the topics studied; and one task for the Task Assistant module, article summarization. In this evaluation, \name\ is configured to use \texttt{gpt-5.2}.

\subsubsection{Topics Studied}
A set of ground-truth topics was defined by two raters and subsequently revised and assigned during the manual analysis of the article collection. This ground truth is used to evaluate the topic modeling module via two distinct analyses: topic generation and topic assignment.

\paragraph{Topic Generation}
Our ground truth consisted of 22 topics. Starting from \emph{Program Repair} as the initial seed, TopicGPT generated 19 topics in its first step, of which 12 matched the ground-truth topics (54\%). Notably, two ground-truth topics (\emph{Code Performance} and \emph{Green Software}) were merged into a single topic (\emph{Software Performance and Energy Efficiency}). In addition, one generated topic, \emph{Automated Test Case Generation}, only partially corresponded to our label \emph{Specification Generation} and was therefore excluded from the identified topics.
During the refinement step, TopicGPT reduced the set to 14 topics and correctly identified 45\% of the ground-truth topics, with the same two mismatches as before. The performance decline is attributable to TopicGPT’s removal of infrequent topics; consequently, software tasks that did not frequently appear, such as \emph{Code Translation}, were discarded.

\paragraph{Topic Assignment}
Comparing the module’s topic assignments with those of the human raters, we obtained a precision of 0.645 and a recall of 0.850. 

An analysis of the topic assignment metrics reveals that the \emph{Benchmarks} label accounted for the highest number of incorrect assignments, with 30 false positives. 47.6\% of the predictions assigned to this label were incorrect. This behavior is explained by the frequent mention of benchmark suites in many papers, even when they do not represent a core contribution, which leads the model to over-assign this topic. Addressing these alone would result in an estimated precision improvement of approximately 9\%.

In contrast, while the \emph{Human–AI Collaboration} label produced fewer false positives in absolute terms, it exhibited the highest misidentification rate (76.9\% of false positives). This trend is attributed to the model’s tendency to assign this label to articles that merely reference human comparisons or interactions, rather than focusing on substantive contributions or studies centered on Human–AI collaboration in software engineering.

\subsubsection{Programming Languages Used}
As before, during the analysis of the papers, the programming languages were manually identified. For this task, the topic modeling module identified 40 programming languages, while the raters identified 47. 
The module achieved an average precision of 0.590 and an average recall of 0.710.

Manual inspection of the assignments revealed that the model frequently tagged Python as one of the languages. This likely stems from the fact that many papers employ Python for data processing or model training, leading to its assignment even when it was not the language under exploration. Additionally, the model often assigned more languages than were actually relevant, suggesting a tendency toward over-assignment and a degree of hallucination.

\subsubsection{Task Assistant}

For the Task Assistant task, we asked the model for a summary of each paper. The summaries were evaluated by two raters according to four parameters on a Likert scale from 1 to 5. The results of the evaluation are presented in Table~\ref{tab:summary-eval}.

Both raters generally agreed the summaries were accurate and free from hallucinations. \emph{Coverage} was the lowest-scoring criterion, as some summaries did not fully capture the most important details; however, this limitation was minor (average score of 4.333). 
The remaining %
criteria received high scores, suggesting that the summaries were typically well-structured, easy to follow, and conveyed key points with minimal redundancy. %
Coherence/Structure had %
the highest standard deviation, indicating that the model’s ability to organize ideas varied the most across summaries.

\begin{table}
\centering
\footnotesize
\caption{Summary Quality Evaluation.}
\label{tab:summary-eval}
\begin{tabular}{lccccc}
\toprule
\textbf{Criterion} & \textbf{Mean} & \textbf{Std} & \textbf{Criterion} & \textbf{Mean} & \textbf{Std} \\ \midrule
Faithfulness & 4.907 & 0.242 & Structure & 4.558 & 0.454 \\
Salience     & 4.333 & 0.367 & Conciseness & 4.648 & 0.334 \\ \bottomrule
\end{tabular}
\end{table}

\subsubsection{Discussion} The results suggest that while human raters cannot be fully replaced during the screening process, employing an LLM as an additional rater can provide valuable supplementary insights, stimulate discussion, and support the refinement of inclusion and exclusion criteria during the snowballing phase.

For more complex tasks such as topic modeling, our findings indicate that current large language models (LLMs) and associated methodologies are not yet sufficiently reliable to operate autonomously. Instead, their effectiveness is maximized in a human-in-the-loop setting, where they function as assistive tools rather than independent decision-makers. In this framework, LLMs can generate an initial set of topics and corresponding assignments, which are then validated and refined by a human reviewer. This approach substantially reduces manual effort while maintaining methodological reliability. Notably, the models did not hallucinate and generated plausible topics, but these often diverged from the task definition. More task-aligned prompt design could improve topic quality and relevance and is left for future work.

In contrast, for less demanding tasks such as summarization, LLMs already demonstrate satisfactory performance, making them well suited for direct use with minimal human intervention, saving time during the article analysis process.

\section{Related work}

Bacinger et al. present a semi-automated system supporting the search and screening phases of literature reviews. It automates the retrieval of articles from multiple sources, helps define search terms, and uses machine learning models to identify relevant papers. It also supports curating and exporting the final set of papers~\cite{bacinger2022system}.
While Bacinger et al.'s tool uses database searches, \name\ uses snowballing, 
as prior studies show it performs as well as or better than database searches~\cite{badampudi2015experiences, jalali2012systematic}.
Moreover, Bacinger et al.'s system does not support data extraction from articles.

Agarwal et al. introduce LitLLM, an LLM-based toolkit for the generation of scientific literature reviews. %
It automatically generates search keywords from user-provided abstracts, retrieves and re-ranks relevant papers, and produces related work text grounded in these papers through Retrieval-Augmented Generation~\cite{agarwal2024litllm}.
Despite its text generation power, LitLLM lacks manual curation or snowballing, unlike ProfOlaf's human-in-the-loop workflow.

He et al. propose PaSa, an LLM-based agent for academic paper search. 
PaSa automates query generation, retrieves and expands results through citation networks, and uses a selector agent to evaluate relevance, thereby enabling comprehensive and accurate literature retrieval~\cite{he2025pasa}. The authors' work emphasizes automated retrieval and screening but does not allow manual curation, whereas \name\ balances automation with the researcher's control.

Li et al. present ChatCite, a tool that automates literature summarization by extracting key elements from papers, generating comparative summaries through an iterative reflective process, and evaluating the results with a novel automatic metric called G-score~\cite{li2025chatcite}. Unlike ProfOlaf, which covers the review cycle from the search to the data extraction phase, ChatCite focuses exclusively on summary generation and is therefore complementary.

\section{Conclusion}
\name\ addresses key challenges in systematic review, combining iterative snowballing with LLM-assisted screening and analysis, striking a balance between automation and human oversight. By improving efficiency, rigor, and reproducibility, it enables researchers to conduct higher-quality reviews with reduced effort. Given the rapidly growing volume of research in software engineering, 
\name\ can be valuable in helping SE researchers keep pace with the field.

\bibliographystyle{ACM-Reference-Format}
\bibliography{sample-base}

\end{document}